\documentclass[preprint,final]{elsarticle}
\usepackage{amsmath}
\usepackage{amsfonts}
\usepackage{lineno,hyperref}
\usepackage[dvipsnames]{xcolor}
\usepackage{subcaption}
\usepackage{xcolor}
\usepackage[utf8x]{inputenc}
\usepackage{chngcntr}
\usepackage{enumitem}
\counterwithout{figure}{subsection}
\usepackage{changes}
\usepackage{mathtools}
\usepackage{siunitx}

\DeclareSIUnit{\Ci}{\text{Ci}}

\journal{Nuclear Inst. and Methods in Physics Research, A}

\makeatletter
\def\ps@pprintTitle{% \let\@oddhead\@empty \let\@evenhead\@empty
 \def\@oddfoot{}%
 \let\@evenfoot\@oddfoot}
\makeatother

\begin{document}

% \includepdf[pages = 1]{coverletter.pdf}

\begin{frontmatter}

\title{Boron Coated Straw-based Neutron Multiplicity Counter for Neutron Interrogation of TRISO Fueled Pebbles}

% % Group authors per affiliation: 
% \author{M. Fang, N. Bartholomew,and A. Di Fulvio\fnref{myfootnote}} 
% \address{Department of Nuclear, Plasma, and Radiological Engineering, \\University of Illinois, Urbana-Champaign}
% \fntext[myfootnote]{Since 1880.}

% or include affiliations in footnotes:
\author[uiuc]{Ming Fang}
\affiliation[uiuc]{organization={Department of Nuclear, Plasma, and Radiological
                        Engineering, University of Illinois at Urbana-Champaign},
                   addressline={104 South Wright Street},
                   city={Urbana},
                   postcode={61801},
                   state={IL},
                   country={United States}}
\author[PTI]{Jeff Lacy}
\author[PTI]{Athanasios Athanasiades}
\affiliation[PTI]{organization={Proportional Technologies Inc.},
                   addressline={12233 Robin Blvd},
                   city={Houston},
                   postcode={77045},
                   state={TX},
                   country={United States}}
                   
\author[uiuc]{Angela Di Fulvio\corref{mycorrespondingauthor}}
\cortext[mycorrespondingauthor]{Corresponding author. Tel.: +1 217 300 3769. Fax.: +1 217 333 2906}
\ead{difulvio@illinois.edu}
%  \affiliation[uiuc]{organization={Department of Nuclear, Plasma, and Radiological
%                         Engineering, University of Illinois at Urbana-Champaign},
%                   addressline={104 South Wright Street},
%                   city={Urbana},
%                   postcode={61801},
%                   state={IL},
%                   country={United States}}

\begin{abstract}
    % \begin{linenumbers}
    Pebble bed reactors (PBRs) can improve the safety and economics of the nuclear energy production. PBRs rely on TRIstructural-ISOtropic (TRISO) fuel pebbles for enhanced fission product retention. Accurate characterization of individual fuel pebbles would enable the validation of computational models, efficient use of TRISO fuel, and improve fuel accountability. We have developed and tested a new neutron multiplicity counter (NMC) based on 192 boron coated straw (BCS) detectors optimized for ${}^{235}$U assay in TRISO fuel. The new design yielded a singles and doubles neutron detection efficiency  of 4.71\% and 0.174\%, respectively, and a die-away time of \SI{16.7}{\micro\second}. The NMC has a low intrinsic gamma-ray detection efficiency of $8.71\times10^{-8}$ at an exposure rate of 80.3 mR/h. In simulation, a high-efficiency version of the NMC encompassing 396 straws was able to estimate the ${}^{235}$U in a pebble with a relative uncertainty and error both below 2\% in \SI{100}{\s}. 

\end{abstract}


\end{frontmatter}

% \linenumbers

\section{Background and Motivation}
Pebble bed reactors (PBRs) are advanced reactors that can potentially improve the safety, efficiency, and economics of the nuclear energy production~\cite{kadak2005future}. PBRs rely on tristructural-isotropic (TRISO) fuel for enhanced fission product retention and improved spent fuel management~\cite{demkowicz2019coated}. The reactor core can contain hundreds of thousands of TRISO-fueled pebbles, which flow continuously through the reactor core and can be reinserted into the reactor until each pebble has reached a target burnup. \replaced{The ability to uniquely identify individual fuel pebbles provides the potential for improvements in both the operational and fundamental research aspects of PBRs. Fuel transit time can be determined more precisely, which can be leveraged to validate computational models. Fuel-use efficiency by means of controlling excessive burnup accumulation or premature fuel discharge can also be improved. Lastly, and perhaps most importantly, fuel accountability in the context of material control and accountability can be enhanced to supplement currently implemented methods.}{Unique identification of individual fuel pebbles would allow determining the fuel transit time for validation of computational models, preventing excessive burnup accumulation or premature fuel discharge, and improving fuel accountability.} One of the signatures for fuel identification is the ${}^{235}$U mass and burnup level, which can be extracted through neutron coincidence counting. In this application, two constraints need to be met: high sensitivity to assay a small ${}^{235}$U mass and rejection of gamma rays to assay ${}^{235}$U and fissile fission products such as ${}^{244}$Cm in partially burned fuel. To meet these constraints, BCS detectors were chosen to build a NMC for interrogation of TRISO-fueled pebbles. BCS, unlike ${}^{3}$He, are readily available and cost-effective. Pulse rise time for BCS and ${}^{3}$He detectors are approximately \SI{100}{\nano\second} and a few \si{\micro\second}~\cite{lacy2009boron},\added{ respectively,} therefore BCS show a faster response than ${}^{3}$He and are less prone to pile up in high count rate scenarios.\added{ The BCS used in this work features a pie-shaped cross section with increased surface area, which provides higher detection efficiency and shorter die-away time than the round straws~\cite{8069783}. The increased efficiency and faster response result in higher assay precision in low-count rate scenarios for fixed dwell times compared to ${}^{3}$He based systems.} BCS detectors are highly insensitive to gamma rays, unlike organic scintillators, hence enabling the assay of irradiated pebbles that have cycled through the core but have not yet reached the end-of-lifetime burnup.\deleted{ The BCS used in this work features a pie-shaped cross section with increased surface area, which provides higher detection efficiency and shorter die-away time than the round straws.}

\added{This paper is organized as follows. In Section~\ref{sec:validation}, a high-fidelity Monte Carlo model of a BCS-based NMC and experiments to validate the Monte Carlo model are described. In Section~\ref{sec:gamma_insensitivity}, the high gamma-ray-insensitivity of the BCS-based NMC are demonstrated. In Section~\ref{sec:rod_interrogation}, active interrogation of uranium samples is carried out using a neutron generator and the BCS-based NMC. In Section~\ref{sec:pebble_interrogation}, the performance of a high-efficiency BCS-based NMC are demonstrated in simulation. Finally, the discussion and conclusions are presented in Section~\ref{sec:conclusion}.}

\section{Experimental Validation of the NMC Model with a ${}^{252}$Cf Source}\label{sec:validation}
\begin{figure}[!htbp]
	\centering
	\includegraphics[width=\linewidth]{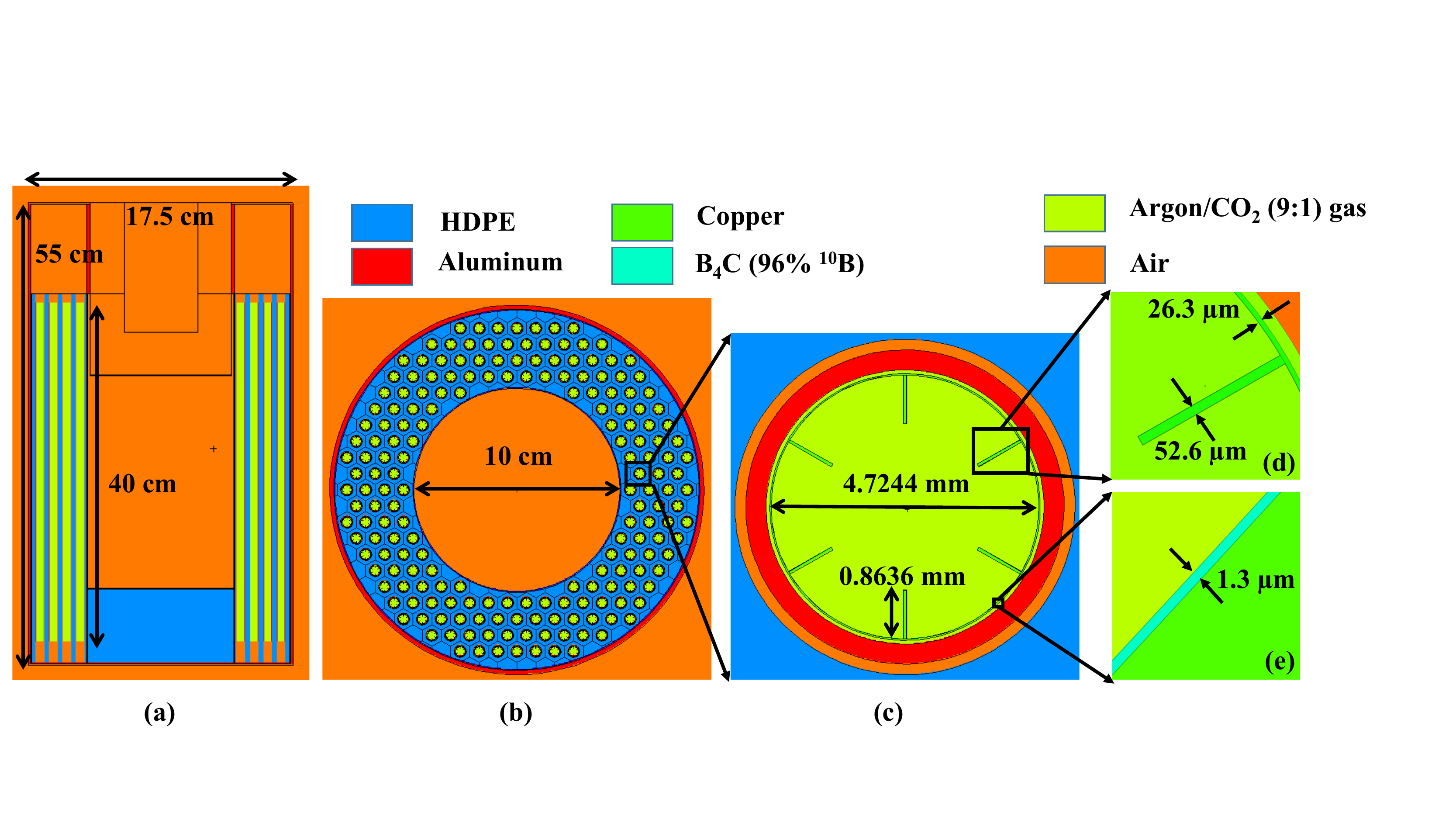}
	\caption{Design of a custom BCS-based NMC for interrogation of TRISO fuel pebbles. (a) is the vertical cross section of the counter, (b) is the horizontal cross section, (c) is the cross section of a pie-6 straw,\replaced{(d) shows the copper septum, and (e) shows }{ and (d) is} the \SI{1.3}{\micro\meter}-thick layer of $\mathrm{{B}_4C}$ \added{coated on the copper substrate }to absorb thermal neutrons.}
	\label{fig:NMC_mcnp_model}
\end{figure}

We have developed a simulation model of a BCS-based NMC in \replaced{Monte Carlo N-Particle (MCNP)}{MCNP} 6.2~\cite{osti_1419730}\deleted{, as shown in Fig.~\ref{fig:NMC_mcnp_model}}.\added{MCNP is a general-purpose Monte Carlo radiation transport code developed and maintained by Los Alamos National Laboratory.}\replaced{ As shown in Fig.~\ref{fig:NMC_mcnp_model}a and \ref{fig:NMC_mcnp_model}b, the NMC is a \SI{55}{\cm}-long and \SI{17.5}{\cm}-diameter cylinder with a central cavity of \SI{10}{\cm} diameter where the sample is placed.}{ The inner and outer diameter of the NMC are 10 cm and 17.5 cm, respectively.}\added{ The geometric efficiency, defined as the percentage of particles emitted by a point-like source in the middle of the cavity that reaches the NMC's inner surface, is 97\%.} The inner and outer surfaces of the NMC are coated with \replaced{\SI{0.508}{\mm}-thick}{0.02"-thick} cadmium to prevent thermal neutrons reentering the sample cavity. The NMC consists of 192 straws arranged in a hexagonal lattice with an inter-distance\added{ between all straws} of \SI{0.9091}{\cm}, surrounded by high density polyethylene (HDPE)\deleted{, as shown in Fig.~\ref{fig:NMC_mcnp_model}a and Fig.~\ref{fig:NMC_mcnp_model}b}. Fig.~\ref{fig:NMC_mcnp_model}c shows the cross section of a single pie-shaped straw. The straw is \SI{40}{\cm} long and has a diameter of \SI{4.7244}{\mm}. A \SI{1.3}{\micro\meter}-thick layer of $\mathrm{{B}_4C}$ (96\% ${}^{10}$B enrichment) is deposited on the inner surface of the straw to absorb the thermal neutrons. There are six septa inside each straw to increase the surface area, thus increasing the detection efficiency and reducing the system die-away time. The straw is filled with Ar/$\mathrm{{CO}_2}$ (9:1) gas mixture at 0.7 atmosphere to detect ${}^{7}$Li/alpha ions released in \replaced{${}^{10}$B(\textsuperscript{1}n,$^4\alpha$)${}^{7}$Li reactions}{${}^{10}$B-neutron capture reactions}. The alpha/${}^{7}$Li particles ionize the gas and the ionized electrons/ions induce electrical signals on the anode wire. In MCNP simulations, we used the PTRAC card to record the energy deposition events in the gas region inside each straw. The timestamps of these events were extracted and analyzed in post-processing.
\begin{figure}[!htbp]
    %\captionsetup{font=footnotesize}
    \centering
    \includegraphics[width=\linewidth]{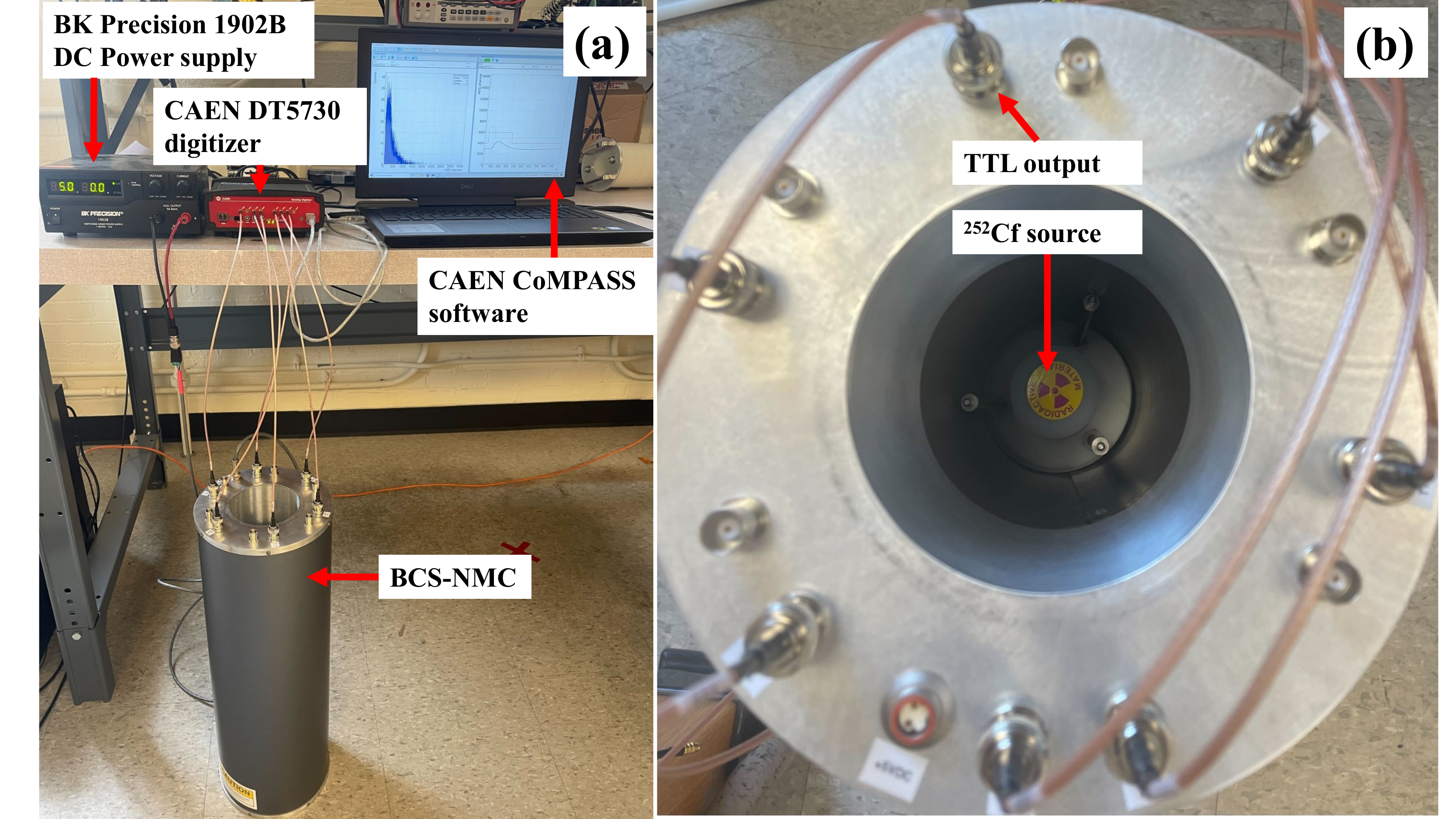}
    \caption{\replaced{$^{252}$Cf experimental setup for validating the MCNP model where the BCS-NMC and all associated electronics are shown in a), and a top view with the \SI{4.943}{\micro\Ci} $^{252}$Cf source placed inside the system cavity is shown in b).}{$^{252}$Cf measurement setup. (a) Overview of the experimental setup. (b) Top view of the NMC with a 4.943-$\mu$ Ci $^{252}$Cf source inside.}}
    % \caption{{$^{252}$Cf experimental setup for validating the MCNP model where the BCS-NMC and all associated electronics are shown in a), and a top view with the \SI{4.943}{\micro\Ci} $^{252}$Cf source placed inside the system cavity is shown in b).}{$^{252}$Cf measurement setup. (a) Overview of the experimental setup. (b) Top view of the NMC with a \SI{4.943}{\micro\Ci} $^{252}$Cf source inside.}
    \label{fig:cf252 measurement}
\end{figure}

A BCS-based NMC based on the design in Fig.~\ref{fig:NMC_mcnp_model} was manufactured by Proportional Technologies (PTI)\deleted{, }. We used the NMC to measure a {\SI{4.943}{\micro\Ci}} ${}^{252}$Cf source\added{ at the date of experiment}, as shown in Fig.~\ref{fig:cf252 measurement}. The ${}^{252}$Cf source placed at the center of the cavity undergoes spontaneous fission with a neutron emission rate of 21,484 n/s. The NMC is powered by a \replaced{low-noise BK Precision 1902B DC power supply}{laboratory power supply} at \SI{5}{\V}, which is converted to a high voltage of \SI{750}{\V} applied to the anode wire of the straws. Signals from 32-straw bundles are summed to one list-mode \replaced{transistor–transistor logic (TTL)}{TTL} output and the six TTL outputs are connected to a CAEN DT5730 digitizer (8 channels, 14 bit, 500 MS/s, 250 MHz bandwidth). The digitizer is controlled by the CAEN CoMPASS software on a laptop for list mode data acquisition~\cite{caencompass}. Additionally, there is an analog output channel connected to one of the six-straw groups, allowing acquisition of analog pulses from the straws and thus the pulse height distribution.

\begin{figure}[!htbp]
        \centering
        \includegraphics[width=\linewidth]{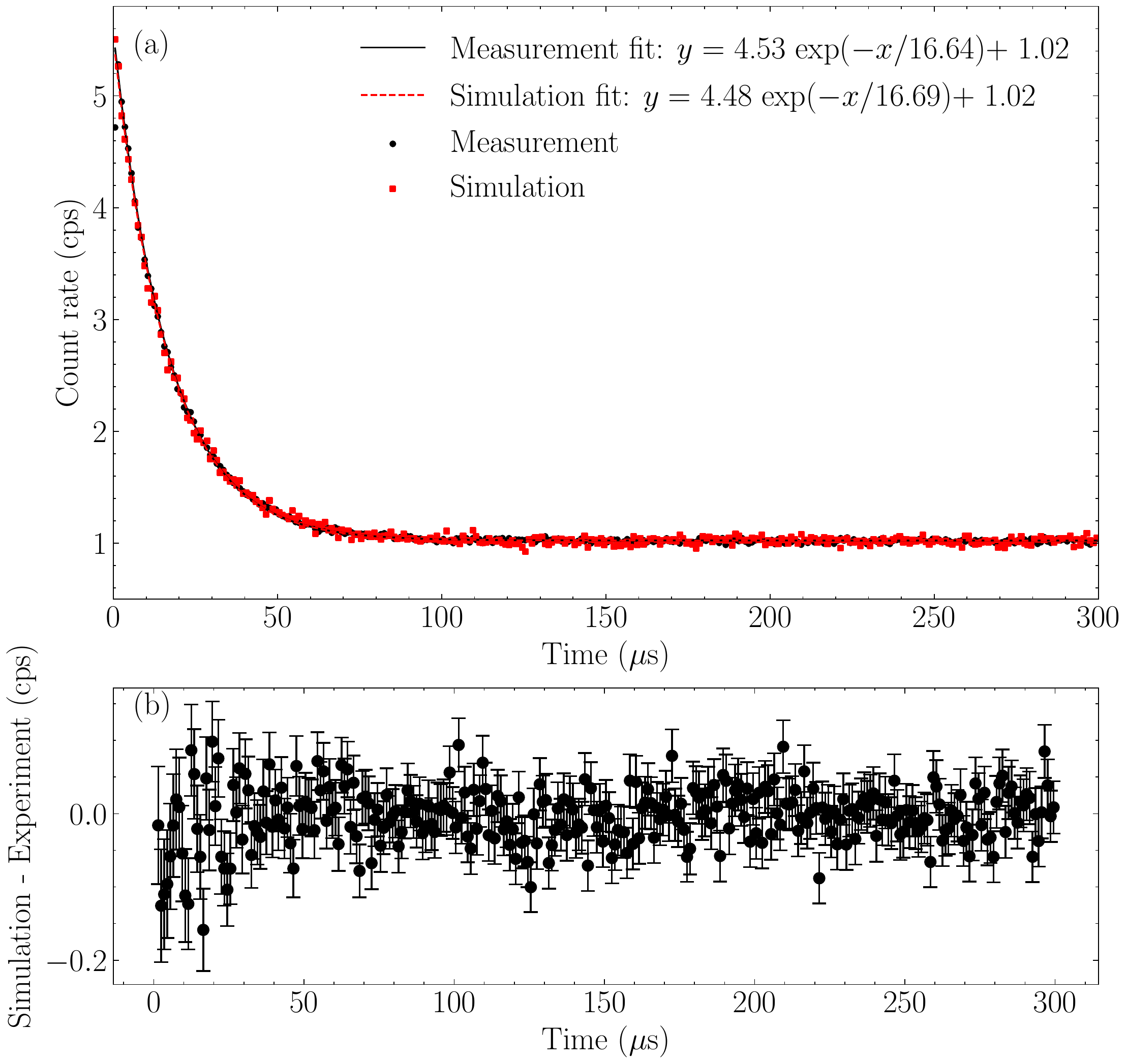}
    \caption{\replaced{Comparison of the experimental and simulated Rossi-alpha distributions, with the distributions and exponential fits shown in a) and the difference between two distributions shown in b). The error bars shown in b) represent 1-sigma uncertainties.}{Top: Comparison of the experimental and simulated Rossi-alpha distributions. Bottom: Difference between the simulated and experimental Rossi-alpha distributions. Measurement time was 3100 s, and simulation time was 1000 s.}}
    \label{fig:ross-alpha}
    % \vspace{-1em}
\end{figure}

To validate the NMC model, we simulated the system die-away time and neutron counts from the ${}^{252}$Cf source in MCNP and compared them to the experimental values. \replaced{The die-away time is a performance parameter of multiplicity counters and measures the mean lifetime of the neutron population in the counter before detection or loss. A short neutron die-away time is desirable to minimize the impact of random coincidences of uncorrelated events and improve measurement precision~\cite{croft2014optimum}. The die-away time can be measured through the analysis of the Rossi-alpha distribution. The Rossi-alpha distribution is the frequency of occurrence of the time intervals between pulses calculated using the time stamps of the list-mode detection events~\cite{hansen1985rossi}. First, we obtained the experimental and simulated Rossi-alpha distributions by binning the arrival time differences between two events according to the standard signal-triggered shift-register scheme~\cite{Croft2012152}, as shown in Fig.~\ref{fig:ross-alpha}. The system die-away time was then obtained by fitting the Rossi-alpha distribution to an exponential function~\cite{orndoff1957prompt,alma99141272912205899,Ensslin20076PN}. The experimental and simulated Rossi-alpha distributions are in good agreement (Fig.~\ref{fig:ross-alpha}), and the relative difference between the experimental and simulated die-away time is 0.3\%.}{We first created the experimental and simulated Rossi-alpha distributions, shown in Fig.~\ref{fig:ross-alpha}. The system die-away time was obtained by fitting an exponential function to the Rossi-alpha distribution~\cite{Ensslin20076PN}. The two distributions are in good agreement, with a relative difference of 0.3\%.} We calculated the neutron count rates using the signal-triggered shift-register algorithm~\cite{Croft2012152}. The pre-delay time was set to \SI{2}{\micro\second}, the gate was set to \SI{32}{\micro\second}, and the long delay time was set to \SI{2}{\ms}.\added{ The uncertainty associated with the singles and doubles were calculated using~\cite{prasad2018analytical}:}
\begin{equation} \label{eq:S_D_uncertainty}
\begin{split}
    \sigma_S &= \frac{S+2Df_1}{T},\\
    \sigma_D &= \frac{D+2S^2G+4DSGf_1+2D^3f_1/S^2}{T}
\end{split}
\end{equation}
\added{where {$S$ and $D$ are the singles and doubles count rates, respectively}, $G$ is the time gate width, $f_1$ is the gate utilization factor, and $T$ is the interrogation time, which was \SI{3100}{\s} for measurement and \SI{1000}{\s} for simulation. } Table~\ref{table:sdt-comparison} shows the comparison of the experimental and simulated die-away time\replaced{ and singles and doubles}{ and coincidence} rates. The relative differences are all below 0.4\%, demonstrating the \replaced{high fidelity}{high-fidelity} of the NMC model.

\begin{table}[!htbp]
    \centering
    \caption{Comparison of the experimental and simulated singles (S), doubles (D) count rates, and die-away time ($\tau$).}\label{table:sdt-comparison}
	\begin{tabular}{l|ccc}
		\hline
		& S (cps) & D (cps) & $\tau$ (\si{\micro\second}) \\ \hline
		Measurement              & $1009.8\pm0.5$  & $59.5\pm0.2$    & $16.64\pm0.06$             \\
		Simulation               & $1008.9\pm1.1$  & $59.3\pm0.4$    & $16.69\pm0.10$             \\
		Relative difference (\%) & \textbf{0.09}    & \textbf{0.34}    & \textbf{0.30}               \\ \hline
	\end{tabular}
\end{table}

\section{Demonstration of High \replaced{Gamma-Ray-Insensitivity}{Gamma Insensitivity} of BCS-based NMC}\label{sec:gamma_insensitivity}
\begin{figure}[!htbp]
    \centering
     \includegraphics[width=\linewidth]{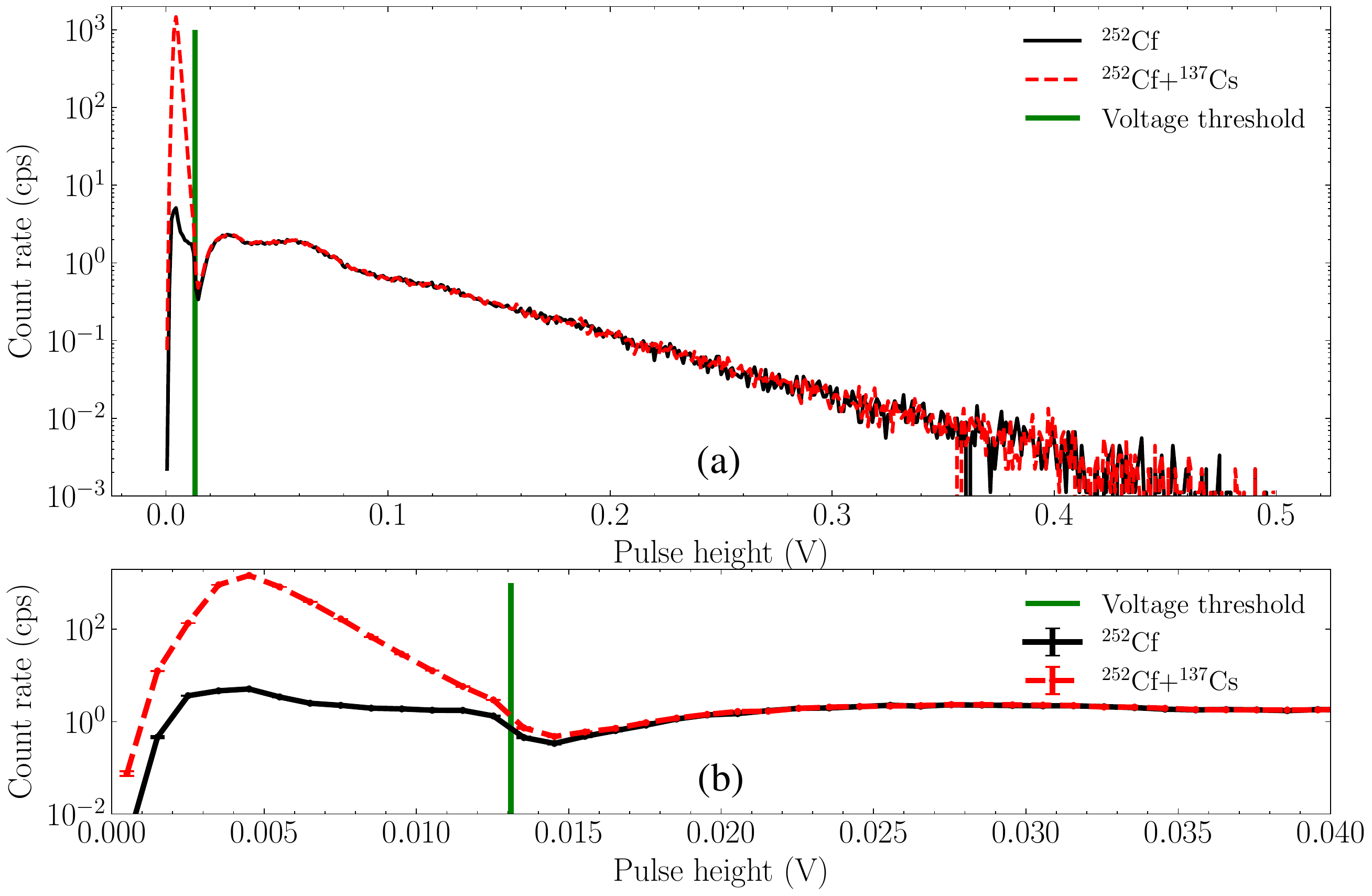}
    \caption{Comparison of the measured ${}^{252}$Cf pulse height spectra from the analog channel with and without the presence of a strong ${}^{137}$Cs source\replaced{, with the full spectra shown in a) and the region near the voltage threshold shown in b). The error bars in b) represent the 1\--sigma uncertainties associated with the count rates, which are difficult to see because the uncertainties are small.}{. Errors are shown as shaded regions.}}
    \label{fig:Cf-CfandCs}
\end{figure}

\begin{figure}[!htbp]
    \centering
     \includegraphics[width=.9\linewidth]{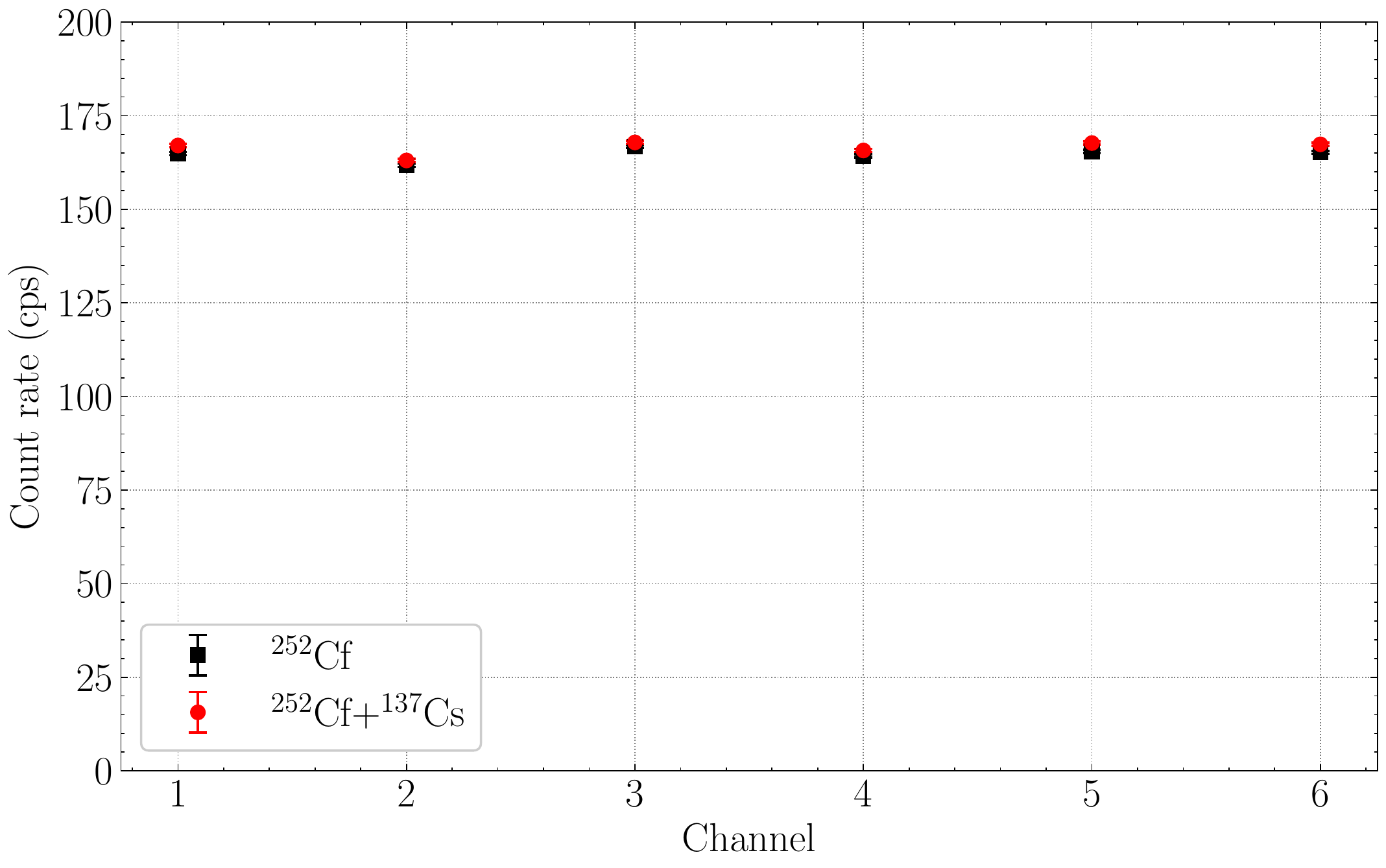}
    \caption{Comparison of the counts in six TTL channels with and without the presence of a strong ${}^{137}$Cs source. \replaced{The error bars shown here represent 1-sigma uncertainties, and can appear smaller than the marker.}{Errors are shown as error bars.}}
    \label{fig:Cf-CfandCs_TTL}
\end{figure}

Similar to BF${}_3$ tubes, \replaced{gamma-ray}{gamma ray} interactions and \replaced{electronic noise}{electronic noises} result in low amplitude events that can be easily rejected by applying a proper voltage threshold~\cite{knoll2010radiation}. We measured two different source configurations to demonstrate the high \replaced{gamma-ray-insensitivity}{gamma ray insensitivity} of the counter: in the first case, a {\SI{4.562}{\micro\Ci}} ${}^{252}$Cf source was placed at the center of the sample cavity; in the second case, a {\SI{2.618}{\milli\Ci}} ${}^{137}$Cs source was added. The gamma-neutron ratio in the second configuration was approximately {4200}:1. Both measurements lasted for 900 seconds. \replaced{Figure}{Fig.}~\ref{fig:Cf-CfandCs} shows the comparison of the two pulse height spectra from the analog channel. The voltage threshold was set to \SI{13}{\milli\volt} by the manufacturer. The counts below the voltage threshold increased by approximately a factor of 100 due to the addition of the strong ${}^{137}$Cs source, while no significant difference between the two spectra was observed above this threshold.
The \replaced{gamma-ray}{gamma} misclassification ratio, defined as the fraction of \replaced{gamma-ray}{gamma} events above the discrimination threshold, is an important parameter for assessing the gamma-neutron discrimination capability of the counter. \replaced{Figure}{Fig.}~\ref{fig:Cf-CfandCs_TTL} shows the comparison of count rates in the six TTL channels \replaced{with and without}{without and with} the presence of the strong ${}^{137}$Cs source. The total count rates above threshold \replaced{with and without}{without and with} the ${}^{137}$Cs source are \replaced{$998.7\pm1.0$~cps and $988.3\pm1.0$~cps}{$988.3\pm1.0$~cps and $998.7\pm1.0$~cps}, respectively. By calculating the relative difference between two count rates, we found a gamma misclassification ratio of $1.04\pm0.15$\%, at the gamma-neutron ratio of {4200}:1.
{The gamma absolute rejection ratio for neutrons (GARRn) metric, defined as the ratio of the absolute neutron detection efficiencies with and without the presence of the strong gamma-ray source, was $1.011\pm0.001$. Therefore the counter meets the gamma-ray discrimination requirement of $0.9<$ GARRn $<1.1$ for neutron detection systems~\cite{kouzes2011neutron}.}

{We placed the \SI{2.618}{\milli\Ci} ${}^{137}$Cs source at the center of the sample cavity and measured the counts for one hour to determine the intrinsic \replaced{gamma-ray}{gamma} detection efficiency of the counter.\deleted{ The fractional solid angle subtended by the source to the counter was 97\%.}  The average exposure rate in the straws during measurement was \SI{80.3}{\milli R\per \hour}. The net \replaced{gamma-ray}{gamma ray} count rate after subtracting the background was $6.96\pm0.06$~cps, and the intrinsic \replaced{gamma-ray}{gamma} detection efficiency found was $8.71\pm0.08 \times 10^{-8}$, which is within the generally-accepted limit of $10^{-7}$ at 10~mR/h exposure rate~\cite{kouzes2011neutron}}.

\deleted{The intrinsic {gamma} efficiency can be further reduced to the order of 10\textsuperscript{-10} by increasing the voltage threshold without significantly affecting the neutron detection efficiency~\cite{jeff20220305email}. In the future, we will use the counter to measure a strong gamma ray source of 1Ci activity to demonstrate the high gamma ray insensitivity.}
% \added{The gamma misclassification ratio can be further reduced by increasing the voltage threshold without significantly affecting the neutron detection efficiency. To illustrate this, we varied the voltage threshold in Fig.~\ref{fig:misclass-threshold} and plotted the gamma misclassification ratio and neutron count rate, as shown in Fig.~\ref{fig:misclass-threshold}. The gamma misclassification ratio decreased by approximately 30\% by raising the threshold to \SI{20}{\milli\volt}, while the neutron count rate decreased by only 3\%.}
% \begin{figure}[!htbp]
%     \centering
%      \includegraphics[width=.9\linewidth]{figs/misclass-threshold.pdf}
%     \caption{Gamma misclassification ratio and neutron count rate in the analog channel as a function of threshold. }
%     \label{fig:misclass-threshold}
% \end{figure}

\section{Active Interrogation of Uranium Rods \replaced{with}{With} a DT Neutron Generator}\label{sec:rod_interrogation}

\begin{figure}[!htbp]
    \centering
    \includegraphics[width=.8\linewidth]{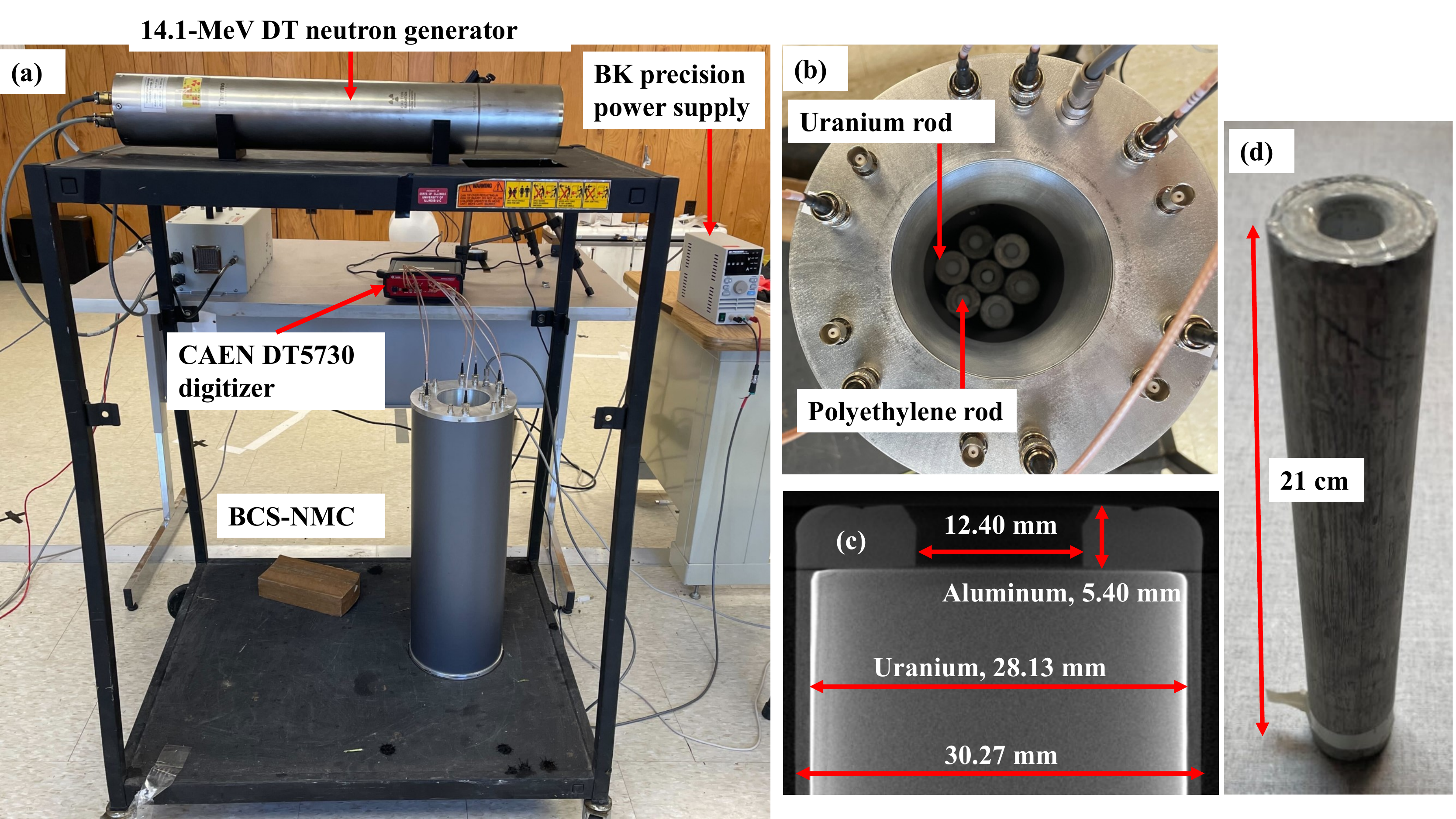}
    \caption{\replaced{Neutron interrogation measurement setup, with the DT generator, NMC and associated electronics shown in a), top view of the counter with the uranium rods inside shown in b), X-ray CT scan of the fuel rod with dimensions labeled in c), and photograph of a single uranium metal rod in d).}{Neutron active interrogation of uranium rods. (a) Overview of the experimental setup. (b) Top view of the NMC with 7 uranium rods inside. (c) Computed tomography cross section of a uranium rod. (d) Side view of a uranium rod.}}
    \label{fig:DTinterrogation_setup}
\end{figure}

We used a \SI{14.1}{\mega\eV} DT neutron generator to induce fissions in a number of uranium rods and measured the \replaced{singles}{single} and \replaced{doubles}{double} count rates using the BCS-based NMC. \replaced{Figure~\ref{fig:DTinterrogation_setup}}{Fig.~\ref{fig:DTinterrogation_setup}} shows the experimental setup.\replaced{ The Thermo Scientific P385 DT neutron generator was operated in continuous mode and the}{ The} voltage and current of the generator were set to \SI{60}{\kV} and \SI{50}{\micro\ampere}, respectively, resulting in a neutron emission rate of $1.75\times10^{7}$ n/s. \added{The TTL output from the DT generator was not used to discriminate between DT neutrons and induced fission neutrons, and both contributed to the measured count rates. }The center of generator target plane was on the axis of the NMC to maximize the number of neutrons reaching the uranium rods. The TTL channels of the NMC were connected to a CAEN DT5730 digitizer, which was controlled by the CAEN CoMPASS software for list mode data acquisition. The uranium metal rod investigated in this study is shown in Fig.~\ref{fig:DTinterrogation_setup}c and Fig.~\ref{fig:DTinterrogation_setup}d. \replaced{Each rod weighs \SI{1.882}{\kg} and is \SI{21.20}{cm} long}{Each rod weights 1.882 kg. The rod is 21.20~cm long} and has an annular cross section with an outer diameter of \SI{3.027}{cm} and an inner diameter of \SI{1.240}{cm}.{ The average enrichment of the rods was 0.049\% $\pm 0.003\%$ determined through \replaced{gamma-ray}{gamma ray} spectroscopy measurements~\cite{reilly1991passive,borrell2009uranium,luca2008experimental}.} We performed an 130~kVp X-ray CT scan of the rod using an Rigaku CT Lab GX130 scanner to determine the thickness of aluminum coating. As shown in Fig.~\ref{fig:DTinterrogation_setup}c, the aluminum coating is \SI{5.4}{mm} thick at the top and bottom surfaces, and \SI{1.07}{mm} thick \replaced{on the side surface of the rod}{elsewhere}. During the measurement, we varied the number of rods from \replaced{zero to six}{0 to 6} and acquired the data for \replaced{30 minutes with four to six rod assemblies and 60 minutes with zero to three rod assemblies to accumulate sufficient amount of counts.}{30-60 minutes for each configuration.}

\begin{figure}[!htbp]
    %\captionsetup{font=footnotesize}
    \begin{subfigure}[t]{0.5\linewidth}
        \centering
        \includegraphics[width=\linewidth]{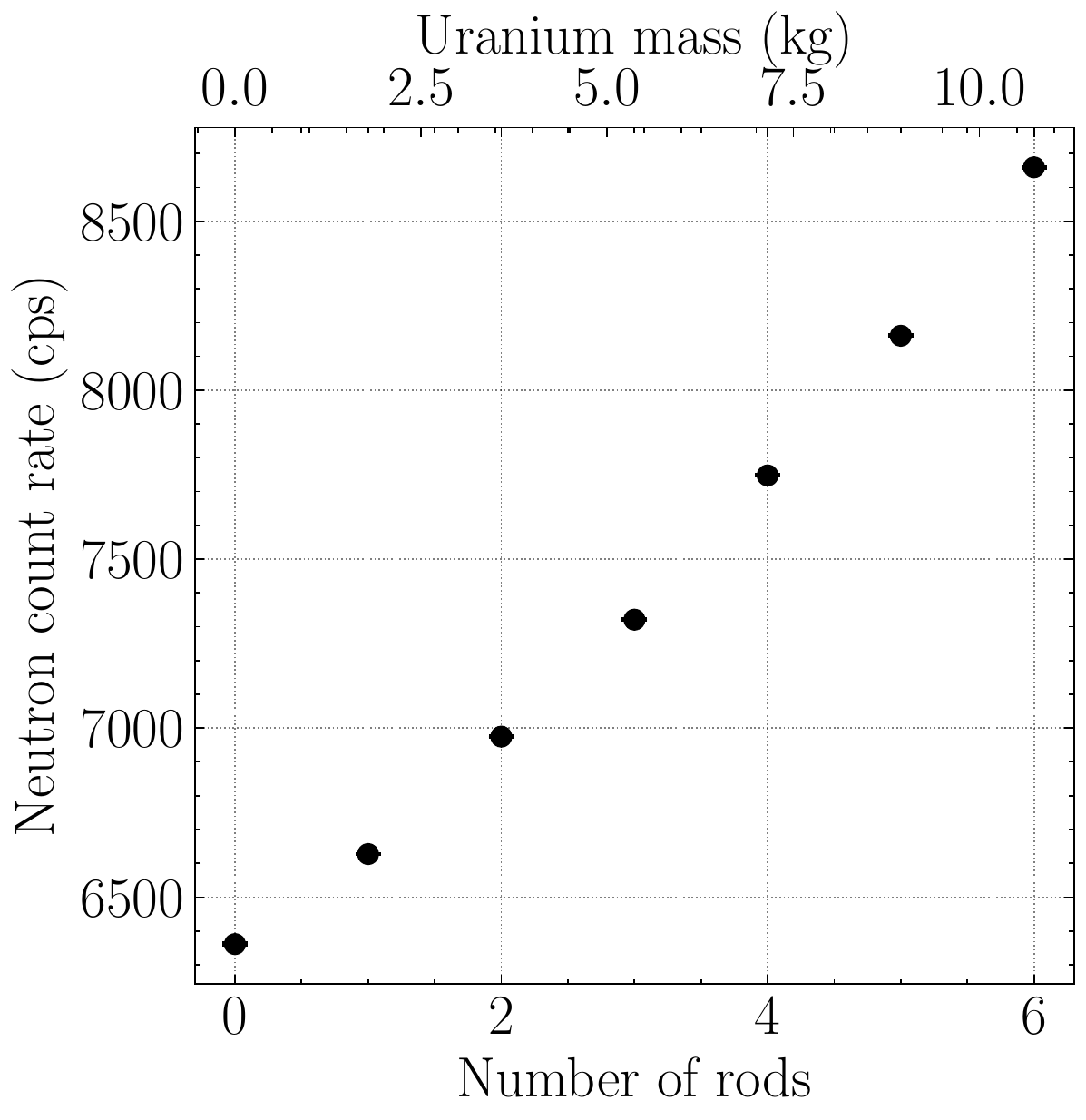}
        \caption{Singles}
        \label{fig:DT_interr_single_exp}
    \end{subfigure}\hfil
    \begin{subfigure}[t]{0.5\linewidth}
        \centering
        \includegraphics[width=\linewidth]{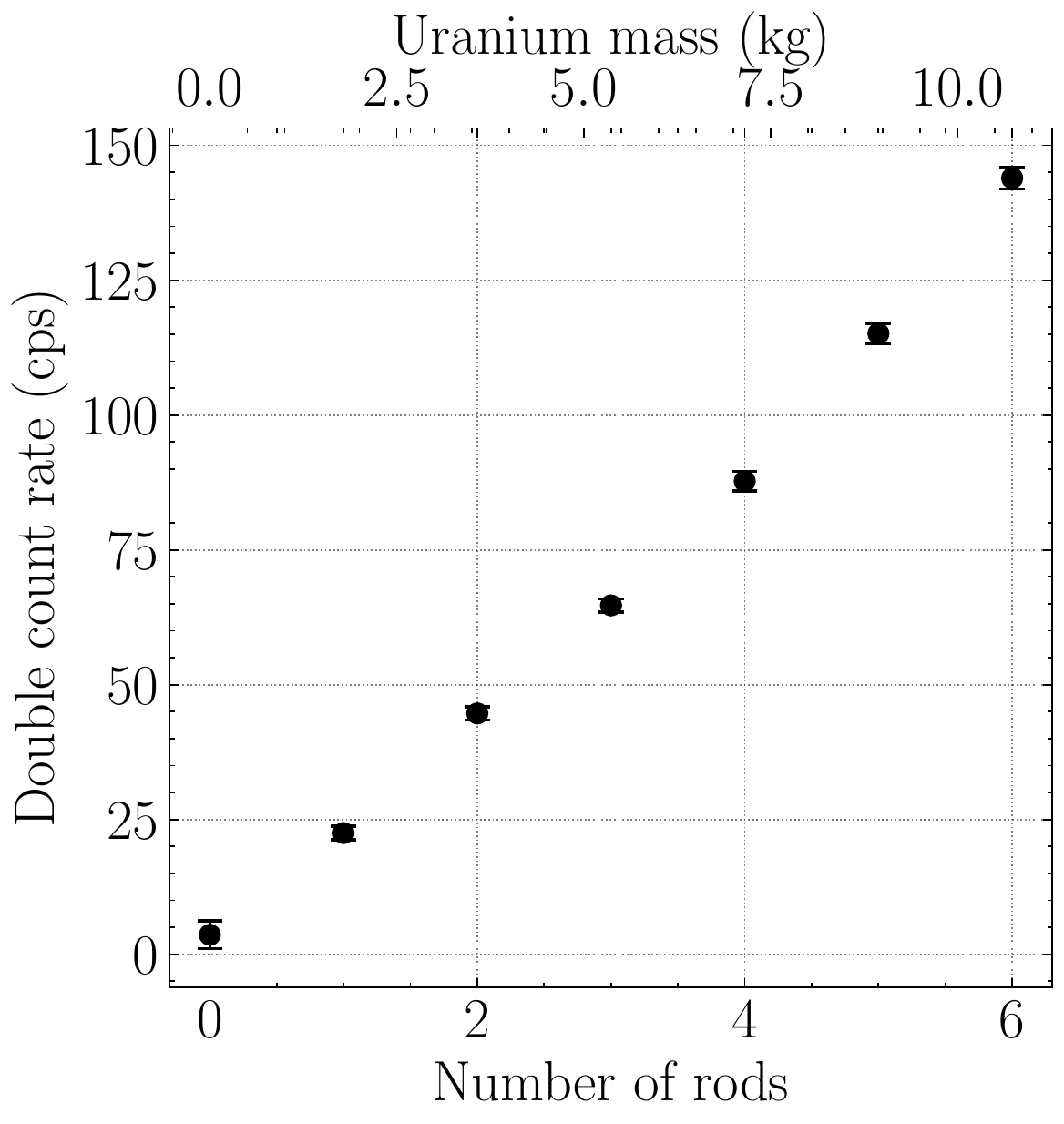}
        \caption{Doubles}
        \label{fig:DT_interr_double_exp}
    \end{subfigure}
    \caption{Neutron singles and doubles rates as a function of number of uranium rods.\added{ The error bars represent 1-sigma uncertainties, and can appear smaller than the marker.}}
    \label{fig:DT_interr_results_exp}
    % \vspace{-1em}
\end{figure}

We calculated the neutron singles and \replaced{doubles}{double} count rates using the signal triggered shift-register algorithm and the results are shown in Fig.~\ref{fig:DT_interr_results_exp}.  \replaced{We used the same parameters as outlined in our validation experiment in section~\ref{sec:validation}, where the}{The} pre-delay time was set to \SI{2}{\micro\second}, the gate was set to \SI{32}{\micro\second}, and the long delay time was set to \SI{2}{\ms}.\added{ Without any rods inside the NMC cavity, i.e., in the zero-rod configuration, the singles and doubles count rates were contributed solely by the DT neutron generator, and the doubles count rate was expected to be near zero since DT reaction does not produce correlated neutrons.} We observed a linear correlation between the uranium sample mass and the measured \replaced{singles and doubles}{single/double} count rates\added{, which is expected since the number of induced fissions is proportional to the amount of fissile material}.\added{ This linear correlation will allow us to estimate the uranium mass in an unknown fuel rod sample by measuring the singles and doubles count rates.}

\section{Simulation of Active Interrogation of TRISO Fuel Pebbles}\label{sec:pebble_interrogation}
\begin{figure}[!htbp]
    \centering
    \includegraphics[width=0.8\linewidth]{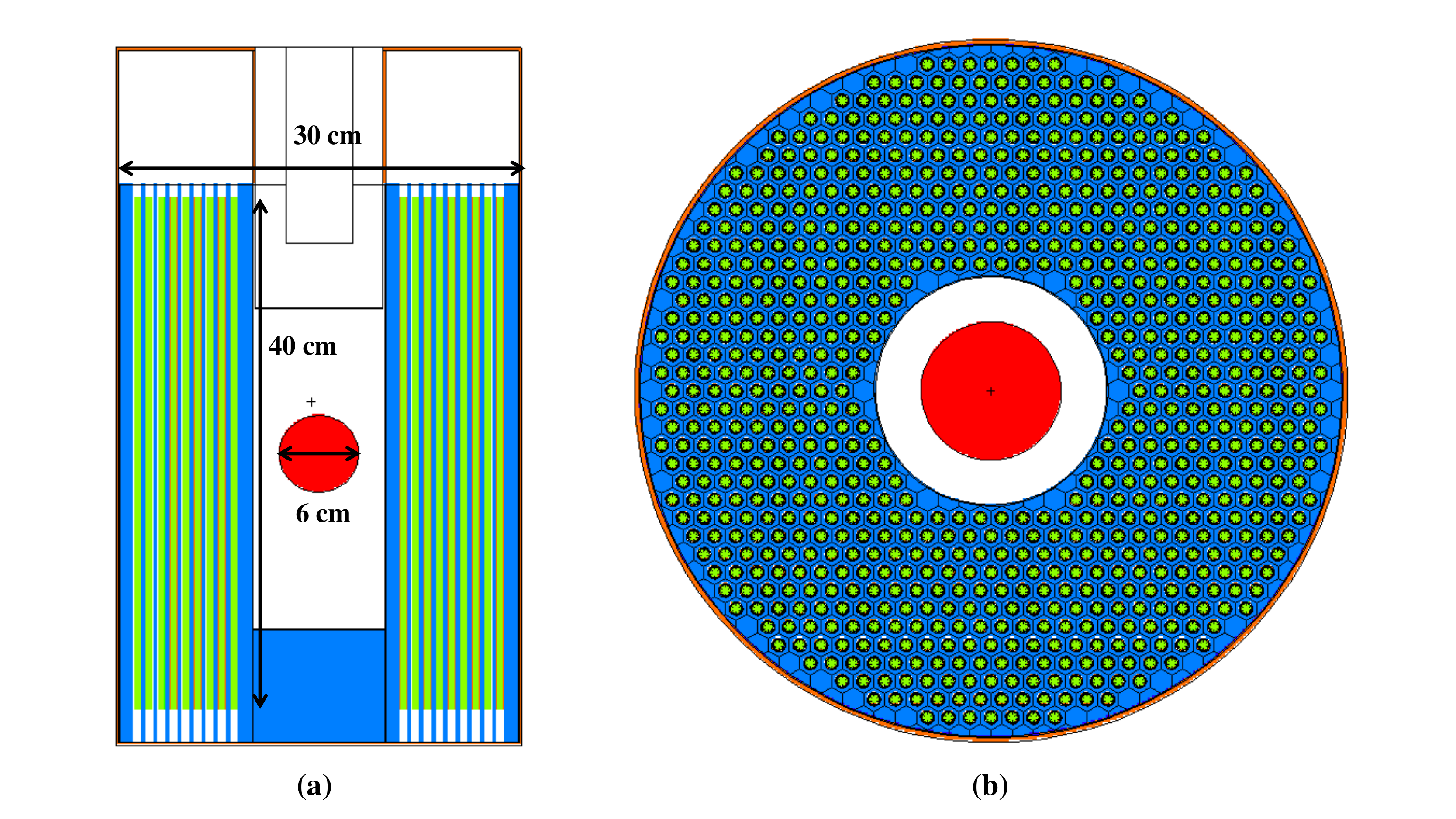}
    \caption{\replaced{The MCNP simulation model showing the 804-straw NMC and a TRISO-fueled pebble inside to be assayed, with the vertical cross section in a) and horizontal cross section in b). A similar model was used to test other configurations, with increasing straw numbers.}{Neutron active interrogation of TRISO fueled-pebbles using an 804-straw NMC.}}
    \label{fig:NMC_804}
\end{figure}

Based on the validated BCS-based NMC model, we have designed a BCS-based NMC encompassing more straws for \replaced{interrogating}{interrogation of} TRISO-fueled pebbles, as shown in Fig.~\ref{fig:NMC_804}. The straws are arranged in a similar way as in our NMC, and the number of straws varies from 192 to 1500. There is a sample cavity of \SI{5}{cm} radius at the center of the counter to accommodate the TRISO-fueled pebble. We simulated the configuration \replaced{with a  ${}^{252}$Cf source placed}{when a ${}^{252}$Cf source was placed} inside the sample cavity to determine the characteristics such as efficiency and die-away time of the new \replaced{BCS-NMCs}{BCS}. The results are summarized in Table.~\ref{table:NMC_characteristics}. The proposed designs achieved shorter die-away time compared to existing round \replaced{BCS- and \textsuperscript{3}He-based}{BCS and \textsuperscript{3}He based} modules.

\begin{table}[!htbp]
\centering
\caption{Characteristics of the NMCs with different number of straws. The existing high level neutron coincidence counter (HLNCC) based on 804 round straws and \textsuperscript{3}He detectors~\cite{simone2017performance} are also shown for comparison.}
\label{table:NMC_characteristics}
\scriptsize
\begin{tabular}{ccccc}
\hline
Model & Diameter (cm) & S efficiency (\%) & D efficiency (\%) & $\tau$ (\si{{\micro\second}}) \\ \hline
192 straws             & 18.0             & 4.71                   & 0.174                  & 16.69              \\
396 straws            & 23.3             & 13.23                  & 1.349                  & 21.56              \\
804 straws           & 30.5             & 23.40                  & 4.175                  & 23.56              \\
1500 straws            & 40.0             & 31.43                  & 7.528                  & 24.36              \\
BCS-based HLNCC~\cite{simone2017performance}             & 34.0             & 13.56                  & N/A                    & 26.00 \\
\textsuperscript{3}He-based HLNCC~\cite{simone2017performance}             & 34.0             & 16.5                  & N/A                    & 43.30 \\
\hline
\end{tabular}
\end{table}

\begin{table}[!htbp]
\centering
\caption{The mass composition of a homogenized pebble of 15\% enrichment. Natural isotopic composition is used if not otherwise specified.}
\label{table:pebble_composition}
\begin{tabular}{|c|c|c|c|c|c|c|}
\hline
Nuclide  & U-235  & U-238  & C        & O      & Si     & Total    \\ \hline
Mass (g) & 1.0128 & 5.7391 & 203.3868 & 0.6264 & 1.6319 & 212.3970 \\ \hline
\end{tabular}
\end{table}

\begin{figure}[!htbp]
    \captionsetup{font=footnotesize}
    \begin{subfigure}[t]{0.5\linewidth}
        \centering
        \includegraphics[width=\linewidth]{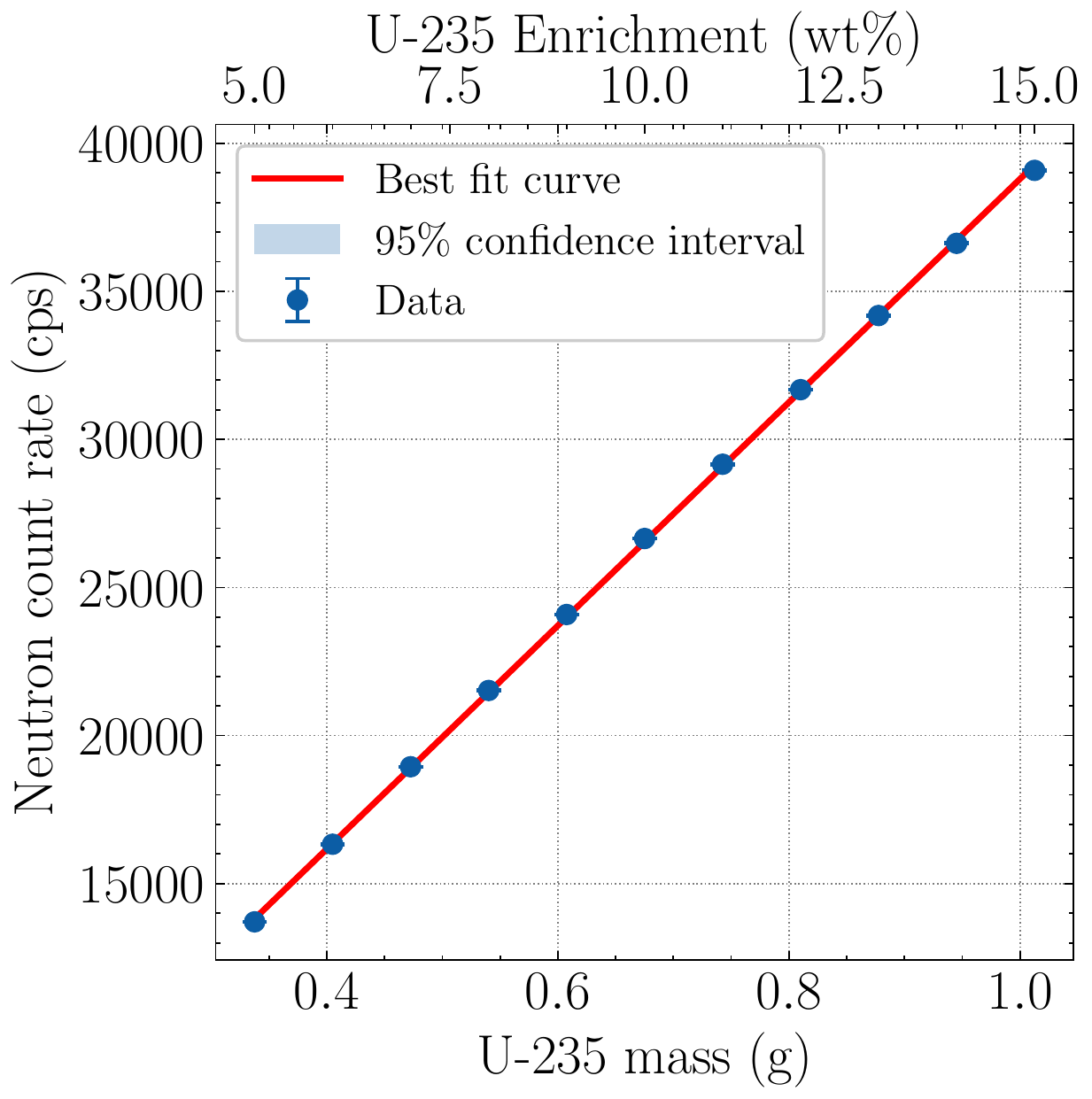}
        \caption{Singles}
        \label{fig:NMC_804_single}
    \end{subfigure}\hfil
    \begin{subfigure}[t]{0.5\linewidth}
        \centering
        \includegraphics[width=\linewidth]{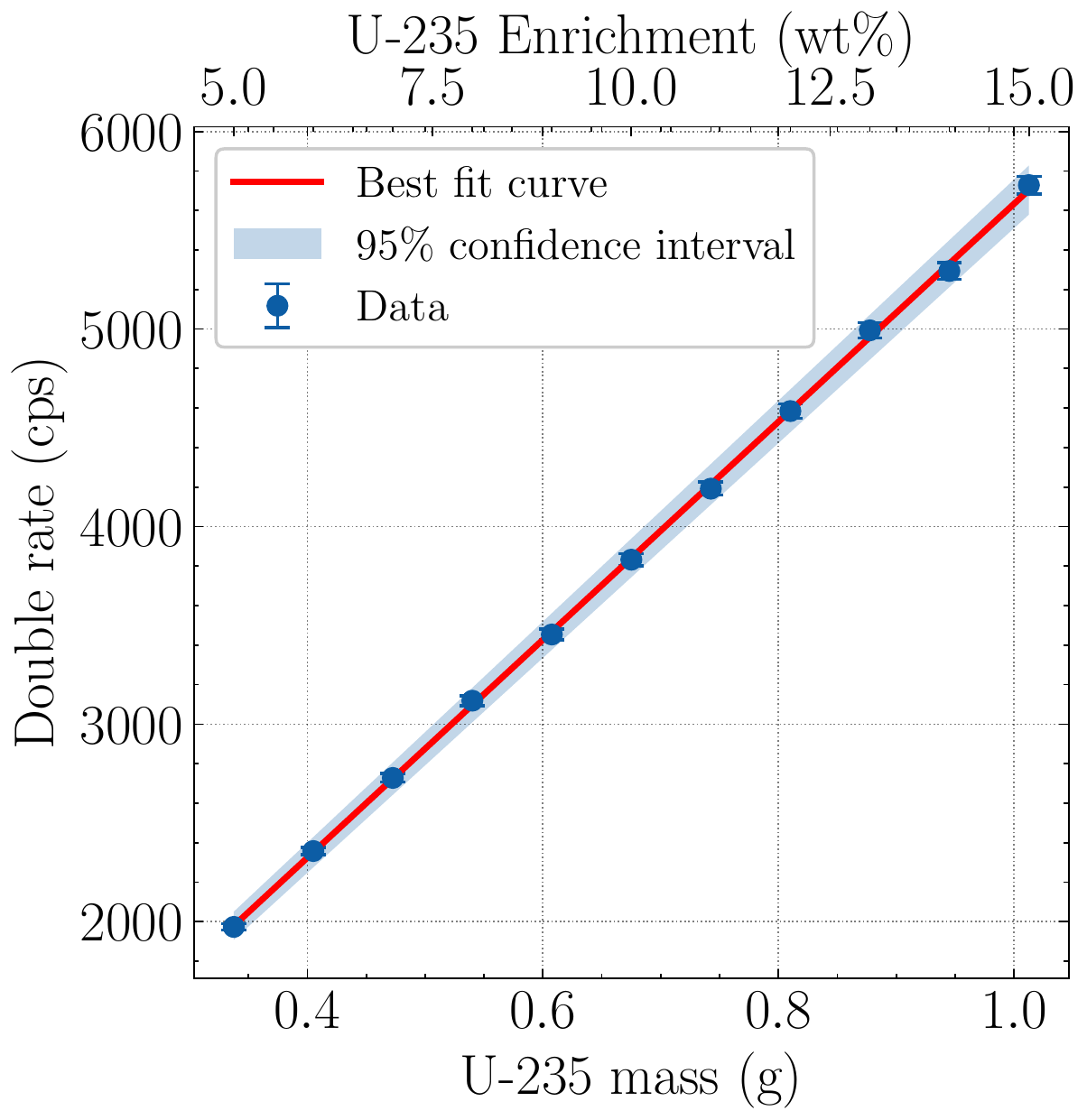}
        \caption{Doubles}
        \label{fig:NMC_804_double}
    \end{subfigure}
    \caption{Neutron singles and doubles rates as a function of ${}^{235}$U mass, obtained with 804 straws.\added{ The error bars represent 1-sigma uncertainties, and can appear smaller than the marker.}}
    \label{fig:NMC_804_results}
    % \vspace{-1em}
\end{figure}

\begin{figure}[!htbp]
    \captionsetup{font=footnotesize}
    \begin{subfigure}[t]{0.5\linewidth}
        \centering
        \includegraphics[width=\linewidth]{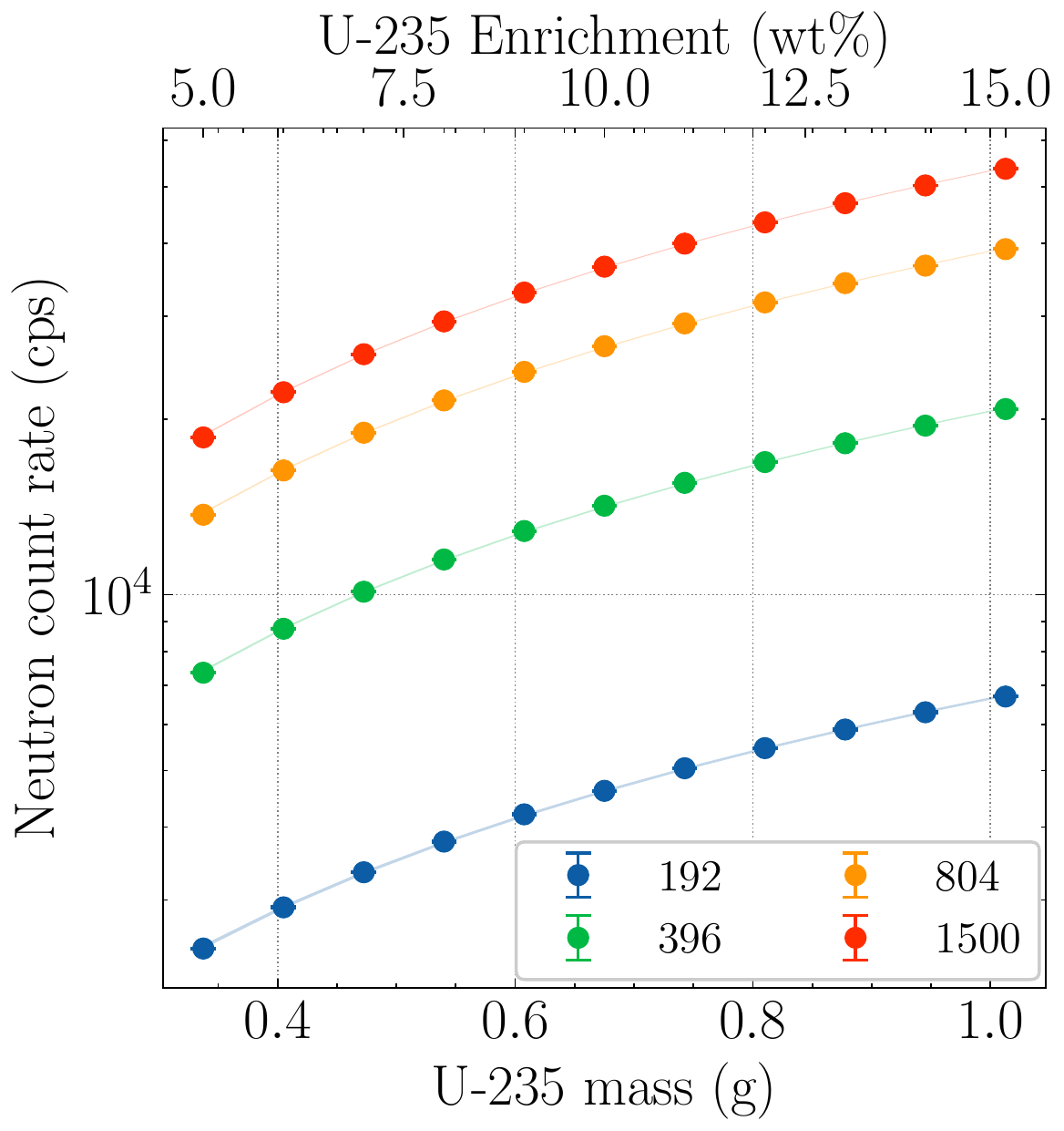}
        \caption{Singles}
        \label{fig:NMC_all_single}
    \end{subfigure}\hfil
    \begin{subfigure}[t]{0.5\linewidth}
        \centering
        \includegraphics[width=\linewidth]{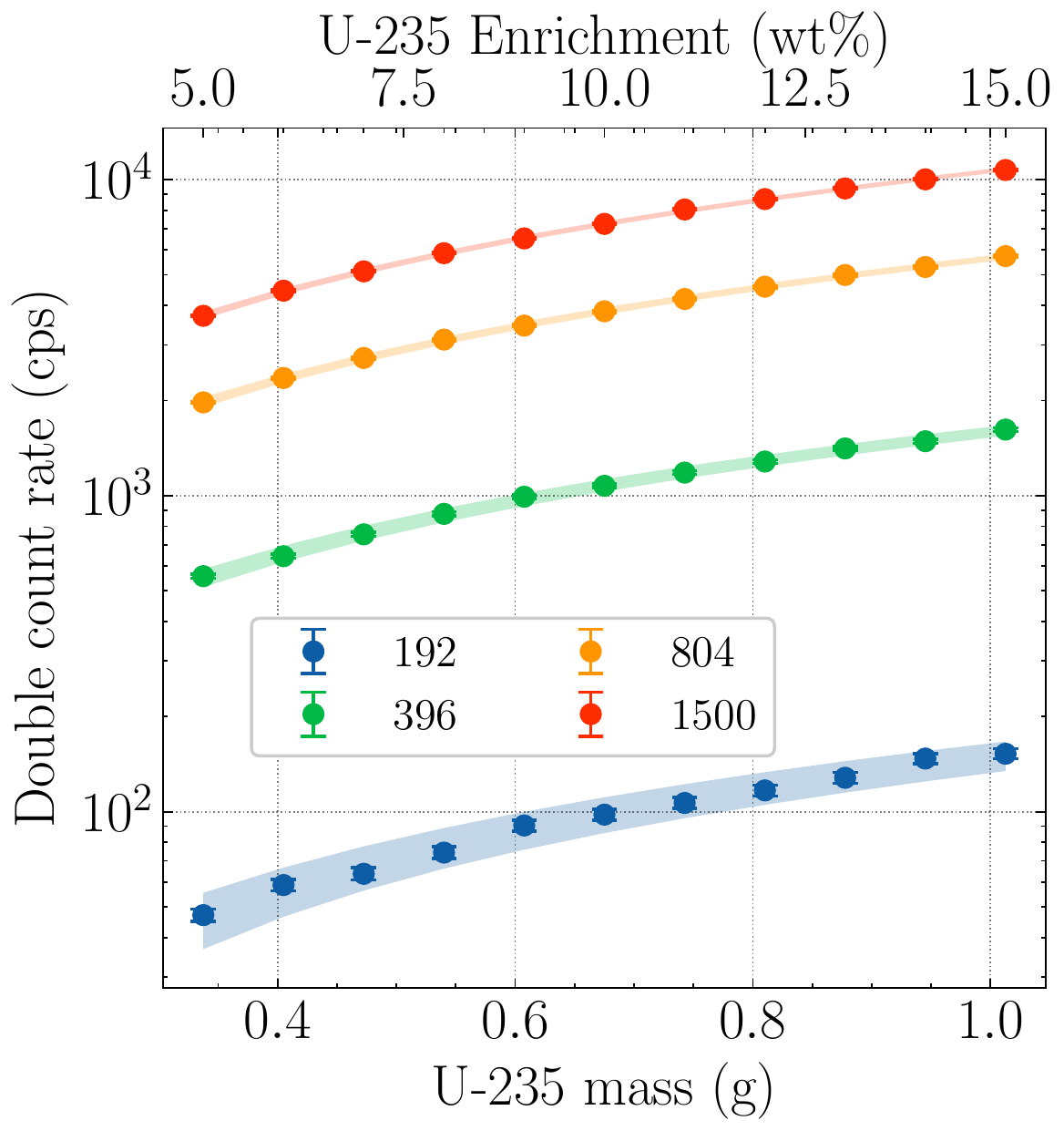}
        \caption{Doubles}
        \label{fig:NMC_all_double}
    \end{subfigure}
    \caption{Neutron singles and doubles rates as a function of ${}^{235}$U mass, obtained with different number of straws.\added{ The error bars represent 1-sigma uncertainties, and can appear smaller than the marker.}\added{ The shaded region represents the 95\% confidence interval of the linear fit.}}
    \label{fig:NMC_all_results}
    % \vspace{-1em}
\end{figure}

We then simulated the active interrogation of TRISO fueled pebbles with enrichment ranging from 5\% to 15\%, as shown in Fig.~\ref{fig:NMC_804}. \added{The TRISO-fueled pebble is a \SI{3}{\cm} radius graphite sphere with 10,000 TRISO fuel particles embedded inside. Each fuel particle contains a UCO fuel kernel of \SI{500}{\um}-diameter surrounded by \SI{100}{\um}-thick buffer, \SI{40}{\um}-thick inner pyrocarbon, \SI{35}{\um}-thick SiC and \SI{40}{\um}-thick outer pyrocarbon. Table~\ref{table:pebble_composition} shows the homogenized fuel material composition for a TRISO-fueled pebble of 15\% $^{235}$U enrichment. }The interrogation source was a thermal neutron beam with a source strength of $10^6$~n/s and the interrogation time was \SI{100}{\s}. We processed the simulation output files and calculated the singles and doubles count rates for the pebbles. \replaced{Figure}{Fig.}~\ref{fig:NMC_804_results} shows the \replaced{singles and doubles}{single/double} count rates as a function of ${}^{235}$U mass. We performed linear fit to both data sets:
\begin{equation}
    y = k x + b\label{eq:linear_fit}
\end{equation}
where $y$ is the \replaced{singles or doubles}{single/double} count rate, $x$ is the ${}^{235}$U mass, $k$ and $b$ are fitting parameters. We repeated the simulation for NMCs with different number of straws and the results are shown in Fig.~\ref{fig:NMC_all_results}.

\begin{figure}[!htbp]
    \centering
    \includegraphics[width=0.7\linewidth]{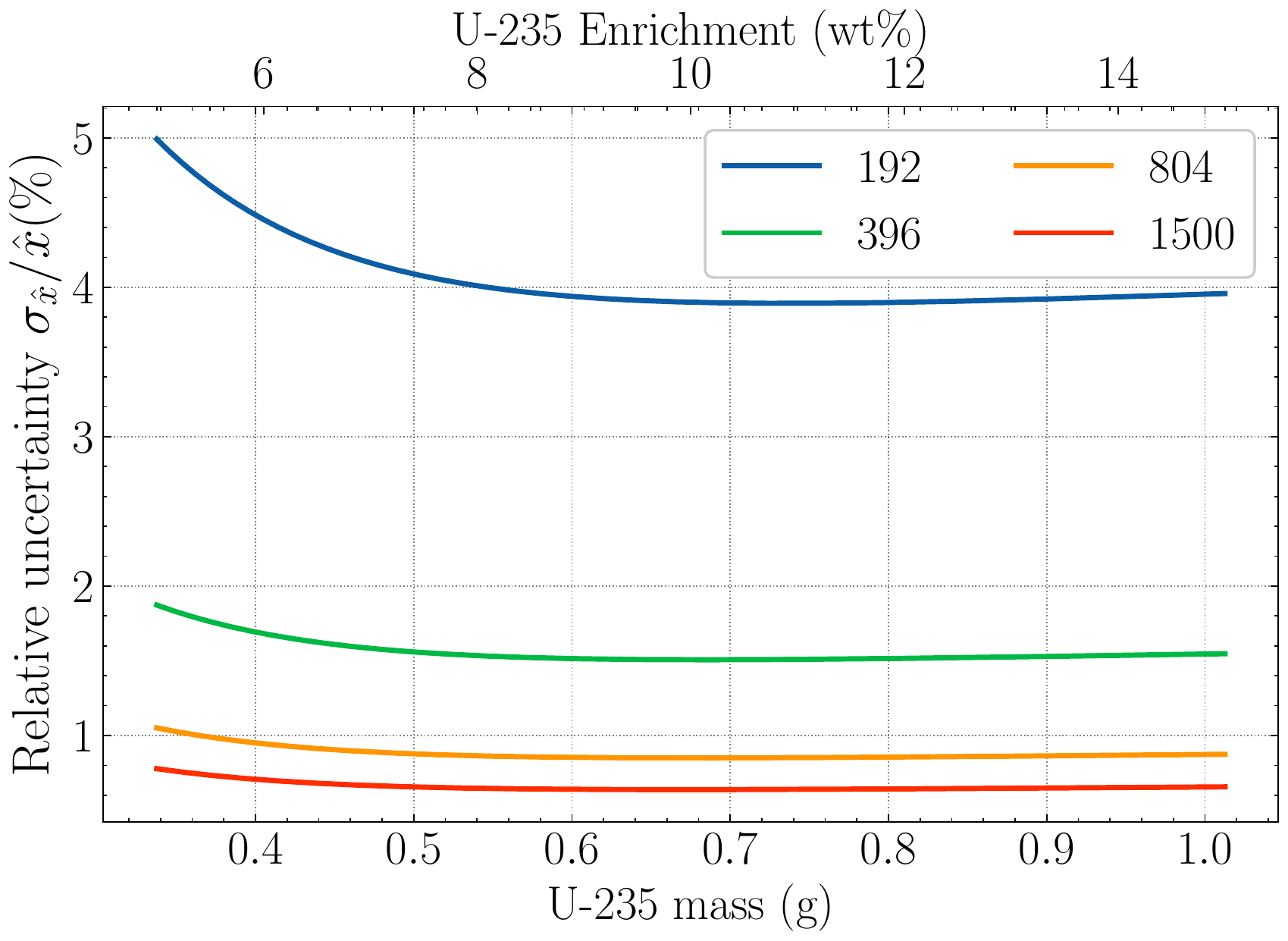}
    \caption{Relative uncertainty associated with $^{235}$U mass estimated based on the double count rate.}
    \label{fig:D_uncerainty}
\end{figure}

\begin{figure}[!htbp]
    \centering
    \includegraphics[width=0.7\linewidth]{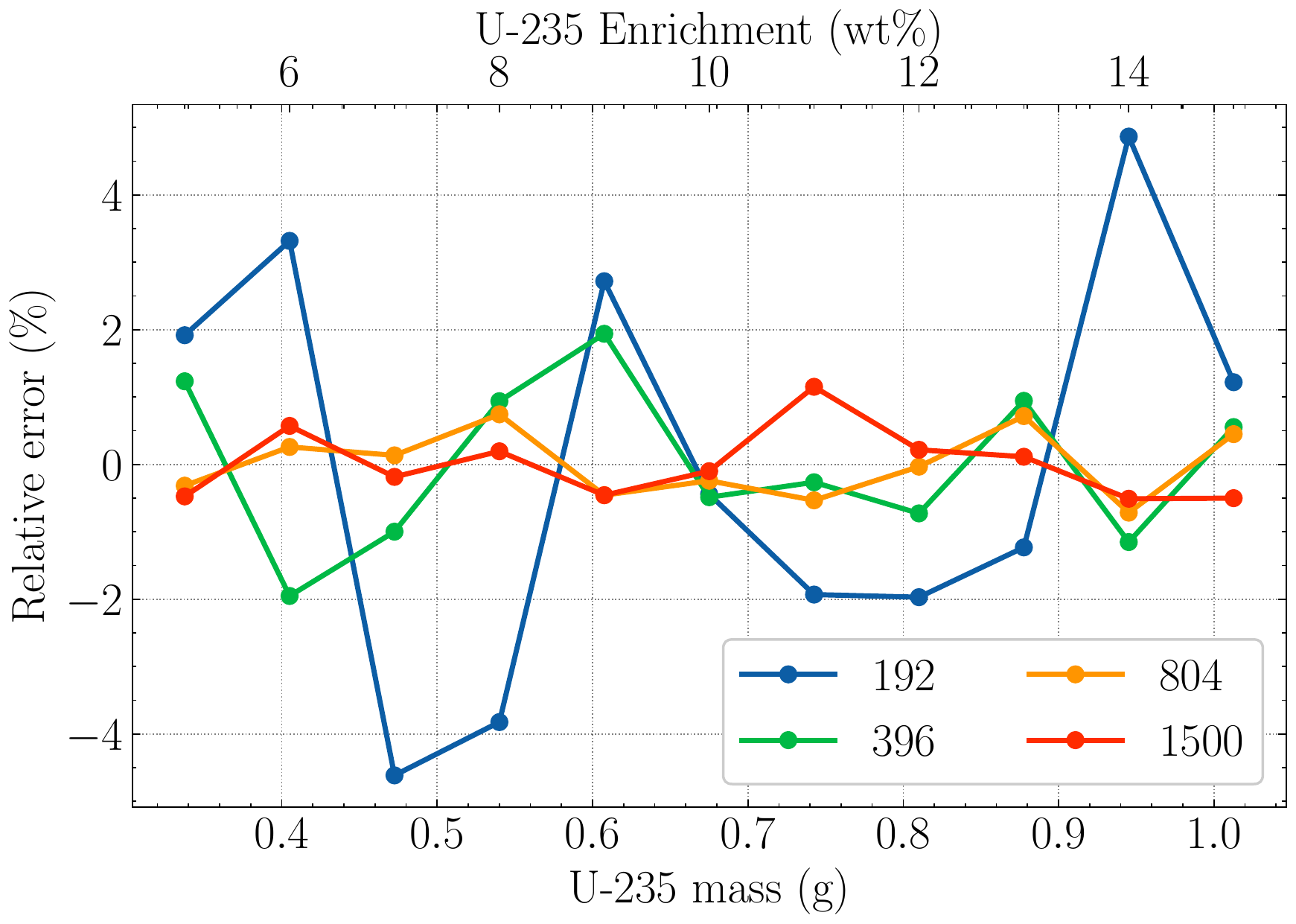}
    \caption{Relative error associated with $^{235}$U mass estimated based on the double count rate.}
    \label{fig:D_error}
\end{figure}

Based on the linear correlation Eq.~\ref{eq:linear_fit}, we can estimate the amount of $^{235}$U $\hat{x}$ in a unknown fuel pebble under inspection by measuring the \replaced{singles and doubles}{single/double} count rates $y$:
\begin{equation}
    \hat{x} = \frac{y-b}{k}
\end{equation}
The uncertainty associated with the mass estimation is:
\begin{equation}
    \sigma_{\hat{x}} = \sqrt{\frac{\sigma_y^2+\sigma_b^2}{k^2}+\hat{x}^2\frac{\sigma_k^2}{k^2}+2\frac{\hat{x}}{k^2}\mathrm{Cov}(k, b)}
\end{equation}
where $\sigma_k, \sigma_b, $ and $\mathrm{Cov}(k, b)$ are the standard deviation of fit parameter $k, b$ and their covariance obtained in the data fitting,  $\sigma_y$ is the uncertainty associated with the measured count rate\added{ shown in Eq.~\eqref{eq:S_D_uncertainty}}\cite{difulvio}.
% \deleted{ For singles, the uncertainty is:
% \begin{equation}
%     \sigma_S = \frac{S+2Df_1}{T}
% \end{equation}
% and for doubles,
% \begin{equation}
%     \sigma_D = \frac{D+2S^2G+4DSGf_1+2D^3f_1/S^2}{T}
% \end{equation}
% where S and D are the singles and doubles count rates, respectively, $G$ is the time gate width, $f_1$ is the gate utilization factor, and $T=$ \SI{100}{\second} is the interrogation time~\cite{prasad2018analytical}.}
We calculated the relative uncertainty associated with $\hat{x}$ for pebbles with enrichment ranging from 5\% to 15\%, as shown in Fig.~\ref{fig:D_uncerainty}. The maximum relative uncertainty occurs at an ${}^{235}$U enrichment of 5\%, due to the lowest count rate.\added{ We also calculated the relative difference between the estimated mass $\hat{x}$ and true mass $x$ to examine the accuracy of ${}^{235}$U mass assay, as shown in Fig.~\ref{fig:D_error}.} We observed that a counter with approximately 400 straws is capable of estimating the ${}^{235}$U mass with a\added{ relative} uncertainty\added{ and error both} below 2\% for all enrichment levels within the 5\%-15\% range in \SI{100}{\second}.

\section{\added{Discussion and }Conclusions}\label{sec:conclusion}
\added{In the worldwide effort to commercialize Generation IV reactor designs by 2030, high-temperature reactors (HTR) with TRISO-based pebbles as fuel are an attractive option because of the inherent safety and high thermal efficiency. PBRs feature continuous refueling by cycling the pebbles through the core. New instruments need to be developed to assay TRISO-fueled pebbles and implement innovative control and fuel management and accountability strategies. 
%Current assessment methods are limited to burnup determination based on gamma-ray spectroscopy, which are challenging at high burnups and are affected by a high measurement uncertainty up to 10\%\cite{hawari2005computational}. 
To address these needs, we have developed a new type of counter based on pie-shaped BCS detectors as an alternative to $^3$He-based counters. Assay of $^{235}$U mass through neutron coincidence counting can provide an independent and accurate measurement of fuel burnup and support combined computational and experimental methods for tracking and identification of TRISO-fueled pebbles in the PBR. %Neutron coincidence counting traditionally relies on the use of $^3$He-based coincidence counters \\
%The BCS-based NMC features superior gamma-ray insensitivity, thermal neutron detection efficiency, and assay accuracy, when compared to multiplicity counters based on $^3$He and organic scintillators.
The active interrogation of uranium rod samples reported in this work demonstrated the NMC's capability to estimate uranium mass through measurements of neutron singles and doubles count rates. Moreover, the BCS-based NMC has a form factor similar to the traditional $^3$He-based counter and can therefore be used as a direct replacement for interrogation of both TRISO-fueled pebbles and conventional fuel rods~\cite{henzlova2015current}.}
\added{A TRISO-fueled pebble contains approximately 1 gram of $^{235}$U and, during operation, the inspection time of each pebble is limited to a few minutes to ensure continuous and uninterrupted pebble re-circulation. 
% For example, the 200~MW\textsubscript{th} Xe-100 reactor has a designed pebble discharge rate of 170 pebbles/day~\cite{xe-100-overview-slides}, and the time interval between the discharges of two pebbles is approximately \SI{500}{\second}. 
Therefore, a highly-efficient NMC is needed to achieve high assay accuracy. Our simulation shows that a BCS-based NMC with 396 straws can estimate the $^{235}$U mass in a TRISO-fueled pebble with a relative uncertainty{ and relative error} both below 2\%, in \SI{100}{\s}.}
%We are currently upgrading our BCS-based NMC from 192 straws to 396 straws to improve the neutron detection efficiency and we will test the upgraded counter on TRISO-fueled compacts.}

\added{The BCS-based NMC is designed for inspection of both fresh and used TRISO-fueled pebbles. A used fuel pebble is a strong gamma-ray emitter with gamma-ray intensity up to $10^{13}$~photons/s~\cite{su2004design}. Therefore, the intrinsic gamma-ray efficiency should be further reduced to the order of 10\textsuperscript{-12}, which can be achieved by increasing the voltage threshold without significantly affecting the neutron detection efficiency~\cite{jeff20220305email}. In the future, we will use the counter to measure a strong source with gamma-ray intensity comparable to used fuel pebble's, such as a \SI{10}{\Ci} \textsuperscript{192}Ir medical source for radiation therapy to demonstrate the gamma-ray insensitivity of the BCS-NMC in a broader gamma-ray flux range.}

\added{In the simulation of active interrogation of TRISO-fueled pebbles, a $10^6$~n/s thermal neutron beam was employed to irradiate the fuel sample and induce fission inside. A high thermal neutron intensity is necessary to achieve the intended assay accuracy within the limited inspection time of \SI{100}{\s}. At a reactor site, a thermal neutron beam line can be constructed to direct the neutrons from the reactor to the fuel sample for interrogation. If a dedicated beam line is not feasible, an alternative option is to use one of the newest-generation neutron generators, which can produce a thermal neutron flux up to $10^7$~n/cm$^2$/s~\cite{dd110m_and_dd109m}.}

To conclude, we have designed and built a NMC based on pie-shaped BCS detectors suitable for interrogation of fresh and partially-spent TRISO fuel pebbles. We validated the MCNP model of the NMC by comparing the simulated and measured die-away time and efficiency, and the relative differences were below 0.4\%. The developed NMC achieved a die-away time of \SI{16.7}{\micro\second}, approximately 61\% lower than $^3$He-based counter and 35\% lower than round BCS-based counter~\cite{simone2017performance}. We demonstrated the high gamma-neutron discrimination capability of the BCS-based NMC {with a \replaced{gamma-ray}{gamma} misclassification ratio of $1.04\pm0.15$\%}, at an impinging gamma-neutron ratio of {4200}:1{, and a \replaced{gamma-ray}{gamma} intrinsic efficiency of  $8.71\pm0.08 \times 10^{-8}$ at a \replaced{gamma-ray}{gamma} exposure rate of {80.3 mR/h}}. We performed active neutron interrogation of uranium rods using a DT neutron generator and the measured neutron count rates were linearly correlated with the uranium sample mass. Finally, based on the validated NMC model, we simulated an extended NMC with 396 straws. The extended NMC is capable of estimating the $^{235}$U mass in a TRISO-fueled pebble with a relative uncertainty\added{ and relative error both} below 2\% in \SI{100}{\s}. We are currently building \replaced{a high-efficiency version of the NMC encompassing 396 straws}{the extended NMC} and in the future we will use it to assay TRISO-fueled compacts. 

\section*{Acknowledgments}
This work was funded in part by the STTR-DOE grant DE-SC0020733. We thank Mr. Alex Sung for his assistance during the active measurement campaign.{ We would also like to thank The Program in Arms Control \& Domestic \& International Security (ACDIS) at UIUC for the ACDIS Summer 2022 Fellowship (to M.~F.).}
\bibliography{references}

\end{document}